\documentclass[12pt]{iopart}

\usepackage[english]{babel}
\usepackage[utf8x]{inputenc}
\usepackage[T1]{fontenc}

\usepackage{iopams}
\usepackage{bm}
\usepackage{mathtools}  

\usepackage{graphicx}
\usepackage{subcaption}
\usepackage{booktabs}
\usepackage{multicol}
\usepackage{placeins}
\usepackage[justification=justified]{caption}

\usepackage{xcolor}
\usepackage{ulem}
\usepackage[colorlinks,allcolors=black]{hyperref}
\usepackage[acronym,nohypertypes={acronym}]{glossaries-prefix}
\usepackage{relsize}
\usepackage[print-unity-mantissa=false]{siunitx}

\newacronym[longplural=amplitude spectral densities]{asd}{ASD}{amplitude spectral density}
\newacronym{aivt}{AIVT}{Assembly Integration Validation Testing}
\newacronym{aei}{AEI}{Albert Einstein Institute}
\newacronym{adc}{ADC}{analog-to-digital convertor}
\newacronym{apc}{APC}{Astroparticule et Cosmologie}
\newacronym{bsim}{BSim}{Beam Simulator}
\newacronym{cic}{CIC}{cascaded integrator-comb}
\newacronym{cte}{CTE}{coefficient of thermal expansion}
\newacronym{cnes}{CNES}{Centre National d'\'{E}tudes Spatiales}

\newacronym{dws}{DWS}{differential wavefront sensing}

\newacronym{esa}{ESA}{European Space Agency}
\newacronym{el}{el}{elliptical}
\newacronym{emc}{EMC}{Electro-Magnetic Compatibility}
\newacronym{eq}{EQ}{equal}
\newacronym{fir}{FIR}{finite impulse response}
\newacronym{fpga}{FPGA}{field programmable gate array}
\newacronym{fmc}{FMC}{FGPA Mezzanine Card}

\newacronym{gw}{GW}{gravitational wave}
\newacronym{grs}{GRS}{gravitational reference sensor}

\newacronym{ifo}{IFO}{interferometers}
\newacronym{ids}{IDS}{Interferometric Detection System}
\newacronym{irfu}{IRFU}{Institut de Recherche sur les lois Fondamentales de l’Univers}
\newacronym[prefixfirst={the}]{lisa}{LISA}{Laser Interferometer Space Antenna}
\newacronym{lr}{lr}{left-right}
\newacronym{lpf}{LPF}{LISA Pathfinder}

\newacronym{lti}{LTI}{linear time-invariant}

\newacronym{lam}{LAM}{Laboratoire d'Astrophysique de Marseille}
\newacronym{lfn}{LFN}{laser frequency noise}
\newacronym{lte}{LTE}{Laboratoire Temps Espace}

\newacronym{mosas}{MOSAs}{Movable Optical Sub-Assemblies}
\newacronym{mosa}{MOSA}{Movable Optical Sub-Assembly}

\newacronym{nu}{nu}{null}
\newacronym[longplural=power spectral densities]{psd}{ PSD}{power spectral density}
\newacronym[longplural=phasemeter]{pm}{PM}{Phasemeters}

\newacronym{pca}{PCA}{Principal Component Analysis}
\newacronym{pr}{PR}{photoreceiver}

\newacronym[longplural=quadrant photoreceivers]{qpr}{QPR}{quadrant photoreceivers}
\newacronym{rin}{RIN}{relative intensity noise}

\newacronym[longplural=interferometers]{ifos}{IFOs}{interferometers}
\newacronym{isi}{ISI}{inter-spacecraft interferometer}
\newacronym{sepr}{SEPR}{single element photoreceivers}

\newacronym{tdi}{TDI}{Time-delay interferometry}
\newacronym{ttl}{TTL}{tilt-to-length}

\newacronym{ud}{ud}{up-down}
\newacronym{uneq}{UNEQ}{unequal}

\newacronym{zifo}{ZIFO}{Zerodur interferometer}

\newcommand{\s}[1]{{\small #1}}

\begin{document}

\title[Validation of optical pathlength stability]{ Validation of optical pathlength stability in a LISA test-bench demonstrator}

\author{Shivani Harer$^{1\star}$, Maxime Vincent$^{1\dagger}$, Hubert Halloin$^{1}$, Ouali Acef$^{8}$, Nisrine Arab$^{2}$, Romain Arguel$^{7}$, Axel Arhancet$^{2}$, Damien Bachet$^{2}$,  Nathalie Besson$^{2}$, Sébastien Bize$^{8}$, Aurélien Boutin$^{9}$, Sara Bruhier$^{5}$, Christelle Buy$^{6}$, Michael Carle$^{4}$, Jean-Pierre Coulon$^{5}$, Nicoleta Dinu-Jaeger$^{5}$, Mathieu Dupont$^{3}$, Christophe Fabron$^{4}$, R\'emi Granelli$^{2}$,David Holleville$^{8}$, Dominique Huet$^{5}$, Pascal Huguet Chantôme$^{10}$, Eric Kajfasz$^{3}$, Mickael Lacroix$^{2}$, Matthieu Laporte$^{1}$, Rodolphe Le Targat$^{8}$, Jean Lesrel$^{1}$, Michel Lintz$^{5}$, Michel Lours$^{8}$, Christophe Meessen$^{3}$, Mourad Merzougui$^{5}$, Alexis Mehlman$^{8}$, Marco Nardello$^{5}$, Laure Oudda$^{7}$, Benjamin Pointard$^{8}$, Pierre Prat$^{1}$, Emmanuelle Rivière$^{7}$, Jérôme Royon$^{3}$, Aurélia Secroun$^{3}$, Samuel Sube$^{2}$, Johannes Veyron$^{10}$, Thomas Zerguerras$^{1}$, Julien Zoubian$^{3}$}
\address{$^1$ Université Paris Cité, CNRS, CEA, Observatoire de Paris – PSL, CNES, AstroParticule \& Cosmologie (UMR 7164), 10 Rue Alice Domon et Léonie Duquet, F‑75013 Paris, France
}

\address{$^2$ CEA, Centre de Saclay, IRFU (Institut de Recherche sur les Lois Fondamentales de l’Univers), F‑91191 Gif‑sur‑Yvette Cedex, France
}

\address{$^3$ Aix‑Marseille Université, CNRS/IN2P3, Centre de Physique des Particules de Marseille (CPPM), F‑13288 Marseille, France
}

\address{$^4$ Aix‑Marseille Université, CNRS, CNES, Laboratoire d’Astrophysique de Marseille (LAM), F‑13013 Marseille, France
}

\address{$^5$ Université Côte d’Azur, CNRS, Observatoire de la Côte d’Azur (OCA), Laboratoire ARTEMIS, 96 boulevard de l’Observatoire, F‑06300 Nice, France
}

\address{$^6$ Université de Toulouse, CNRS/IN2P3, UPS, Laboratoire des 2 Infinis – Toulouse (L2IT), F-31062 Toulouse Cedex 9, France}

\address{$^7$ Centre National d'Études Spatiales (CNES), 18 avenue Edouard Belin, 31400 Toulouse, France}

\address{$^8$ Observatoire de Paris – PSL, CNRS, Sorbonne Université, Laboratoire national de métrologie et d’essais (LNE), Laboratoire Temps‑Espace (LTE, UMR 8255), 61 Av.de l’Observatoire, F‑75014 Paris, France}

\address{$^9$ EXAIL Technologies, Saint Germaine en Laye, France}

\address{$^{10}$ Bertin Technologies, 10 bis avenue Ampère, 78180 Montigny-le-Bretonneux, France}

\ead{harer@apc.in2p3.fr$^{\star}$, mvincent@apc.in2p3.fr$^{\dagger}$.}
\vspace{10pt}
\begin{indented}
\item[]\today
\end{indented}
\newpage
\begin{abstract}
The \gls{lisa} observatory is a future L3 mission of the \gls{esa} to detect gravitational waves, set to launch in 2035. The detector constellation will conduct interferometry to picometer stability over an unprecedented armlength of \qty{2.5}{} million kms. In this paper, we present the development and testing results for the \gls{zifo}, an optical demonstrator built to validate critical technology for the test setup of \gls{lisa}'s interferometric core. Optical pathlength stability measurements on the \gls{zifo} demonstrate successful reduction of bench noise to maintain the \qty{10}{\pico\meter\per\sqrt{\hertz}} specification across the \qty{1}{\milli\hertz} to \qty{1}{\hertz} frequency band. We also identify and characterize dominant noise sources from phasemeters and correlations of beam tilt into the pathlength that are observed during the test campaign.
\end{abstract}

\submitto{\CQG}

\clearpage

\glsresetall

\section{Introduction}
\label{sec:introduction}

The \gls{lisa} is a space-based gravitational wave detector led by the \gls{esa} scheduled to launch in \num{2035}. The detector will use interferometry to measure variations in optical pathlength to picometer stability over an unprecedented detection arm length of 2.5 million kms. Such an arm length allows access to the low frequency band of \qty{0.1}{\milli\Hz} to \qty{1}{\Hz}, populated by information-rich sources like galactic binaries, extreme-mass ratio inspirals and massive black-hole binaries.

LISA is composed of three spacecraft flying in a triangular formation on a heliocentric orbit describing a cartwheel motion, with the orbit trailing 20 degrees behind Earth. The measurement of the variation of optical pathlength will be performed with heterodyne interferometry\footnote{Interferometric measurement with an offset of frequency between the interferometric beams}. The spacecrafts are equipped with two units of the instruments, called \gls{mosa}, to be able to do the measurement on each arms of the triangular interferometer yielding a total of 6 \gls{mosa}s. Each \gls{mosa} is composed of a laser and telescope to receive and send the beam to and from the distant spacecraft, an optical bench and a phasemeter to do the interferometric measurement and a gravitational reference system housing a free falling test mass to act as a geodesic reference.


The \qty{1.5}{\watt} laser beam is transmitted by the spacecraft telescope with a diameter of \qty{30}{\cm}. Due to diffraction-limited divergence, the beam expands during propagation to the distant spacecraft, leading to a kilometer-scale beam waist at the remote spacecraft. The corresponding received power is extremely low, of the order of \qty{700}{\pico\watt}~\cite{LISA:2017pwj}, so the distant spacecraft cannot simply reflect the incoming beam and must operate as an active transponder. A newly emitted beam is phase-locked to the received beam with a frequency offset such that the heterodyne frequency lies between \num{5} and \qty{28}{\mega\hertz}.
 
The interferometric measurement of LISA will be a combination measurement made by three different interferometers; the inter-spacecraft interferometer, the test-mass interferometer and the reference interferometer. The inter-spacecraft interferometer combines the beam from the distant spacecraft with the local reference laser beam. This interferometer will contain the gravitational wave signal, earning it the name of science-interferometer. The test-mass interferometer combines the local laser light reflected from the local test mass with the laser from the adjacent \gls{mosa} to measure the longitudinal distance between the spacecraft and the local geodesic reference. The reference interferometer, on the other hand, combines the local and adjacent laser beams to measure the differential phase noise between the two units of the instrument, providing a common phase reference.

The relative motion between the spacecraft due to the nature of their orbits induces an arm length variation of about 1\%, resulting in unequal arm interferometers that do not suppress \gls{lfn} like ground-based observatories. 
A standard Michelson interferometer set-up overlooking the arm length mismatch would violate the \qty{10}{\pico\meter\per\sqrt{\hertz}} individual noise allocation for LISA by seven orders of magnitude ~\cite{LISA:2017pwj}. 
\gls{tdi} is a post-processing technique that will be used to reduce \gls{lfn} via the construction of virtual equal-arm length interferometers by delaying and linearly combining the split-interferometric measurements ~\cite{Tinto:1999yr,Estabrook:2000ef,Tinto:2020fcc}. The efficiency of \gls{tdi} in reducing laser noise has been extensively studied and experimentally verified for LISA ~\cite{Tinto:2003vj,Shaddock:2003dj} and also other space-based observatories ~\cite{TianQin:2015yph,Hu:2017mde}.
\begin{figure}[h]
  \centering
  \includegraphics[width=0.8\textwidth]{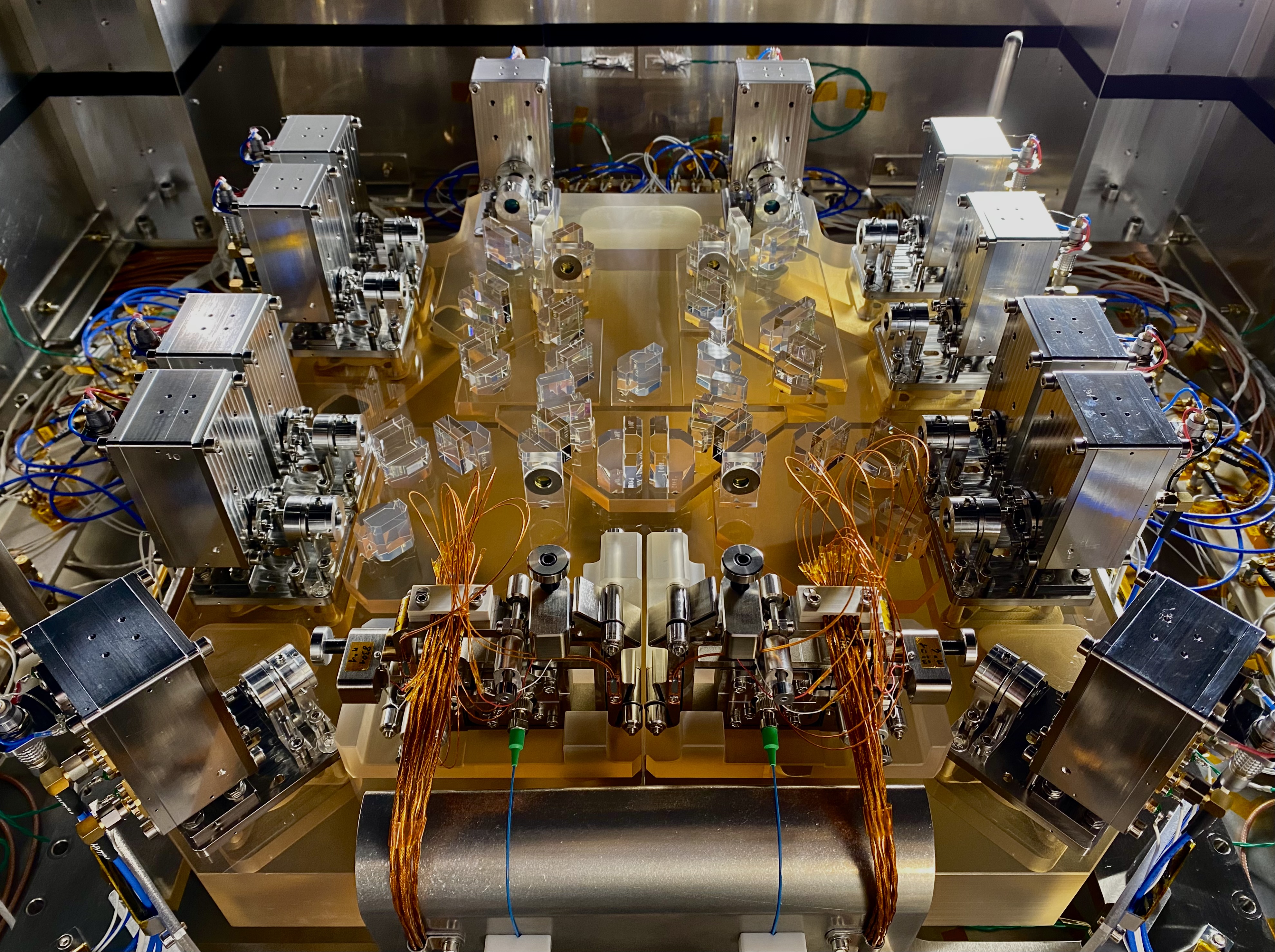}
  \caption{Zerodur Optical Bench placed in its thermal shield in the ERIOS vacuum tank}
  \label{fig:mon_image}
\end{figure}

Apart from \gls{lfn}, LISA sensitivity is broadly limited at low-frequencies of \qty{0.1}{\milli\Hz} to \qty{3}{\milli\Hz} by the residual test-mass acceleration noise and by optical metrology noise at higher frequencies from \qty{3}{\milli\Hz} to \qty{1}{\Hz}. The acceleration noise originating from the differential acceleration of the test masses due to parasitic forces is budgeted at \qty{3}{\femto\meter\s\squared\per\sqrt{\Hz}}. Optical metrology noise allocates budgets for noise contributions from the interferometric measurement system, the optical bench, telescope, phase measurement system, laser and clock system. The proposed threshold for all interferometric-origin noise in a one-way pathlength measurement to be able reach the specified mission performance is \qty{10}{\pico\meter\per\sqrt{\Hz}}~\cite{LISA:2024hlh}.  

The development and testing of the spacecraft is distributed among ESA member states and NASA \cite{LISA:2017pwj}. France's contribution to the instrumental aspect of the project is to perform the full functional and performance testing of the engineering and qualification models of the interferometric core of the instrument. To prepare for this multi-year development and test campaign, French laboratories, in collaboration with the \gls{cnes}, have developed two ultra stable optical benches; the first, a Metallic Interferometer optical bench (MIFO) and a second upgraded Zerodur equivalent, the \gls{zifo}. 

The purpose of these two optical benches was to develop a complex test setup to ensure that every noise source is characterized and fully understood, enabling the successful measurement of optical metrology noise at \qty{10}{\pico\meter\per\sqrt{\hertz}}, in accordance with the requirement for achieving full mission performance. The industrial partner (Bertin-Winlight) responsible for the fabrication of the \gls{zifo} is now contracted for the optical test bench being developed for the next phase of testing, paving the way for the full functional and performance verification of LISA. 

In this paper, we will present the results of the testing campaign of the \gls{zifo}, focusing on the individual contributions of optical metrology noise to the optical pathlength stability measurement. The structure of the paper is as follows. In Section \ref{sec:architecture}, we describe the architecture of the optical bench and the experimental setup. Section \ref{sec:acquisition} describes the data acquisition and processing procedure. Finally, in Section \ref{sec:performanceandnoisereduction} we discuss the results of the test campaign where we validate the optical pathlength stability performance and review different noise sources against the instrument performance model. Concluding remarks are presented in Section \ref{sec:conclusion}.

\section{Instrument architecture }
\label{sec:architecture}

\begin{figure}[h]
    \centering
    \includegraphics[width=\linewidth]{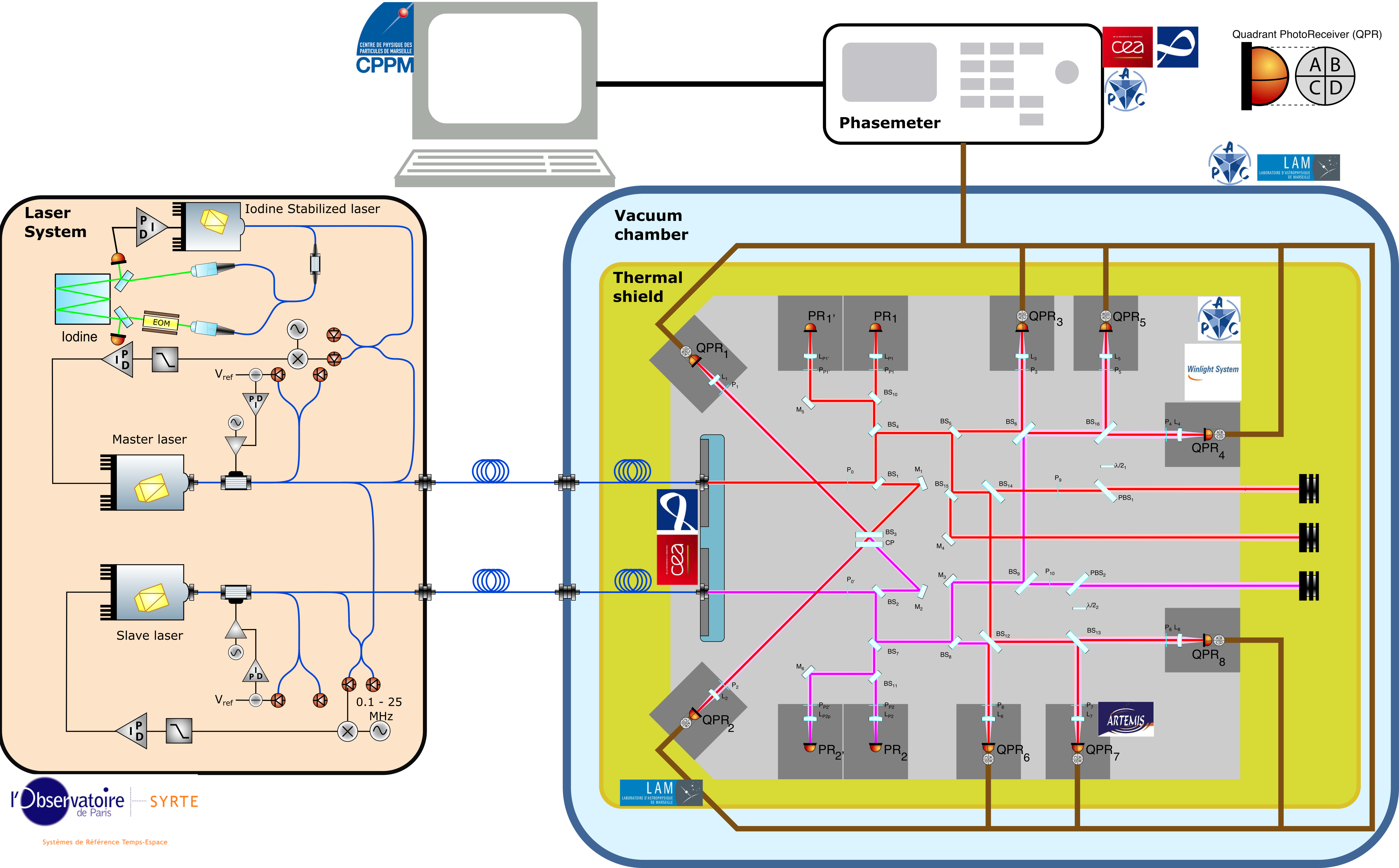}
    \caption{Schematic representation of the \gls{zifo} interferometry bench and entire test set-up along with corresponding partner.}
    \label{fig:instrument design}
\end{figure}

\begin{table}[h]
\centering
\renewcommand{\arraystretch}{1.3} 
\setlength{\tabcolsep}{10pt} 
\begin{tabular}{@{}c l l l@{}} 
    \toprule
    \textbf{IFO no.} & \textbf{Armlength difference $\Delta L$} & \textbf{QPR} & \textbf{PM} \\ 
    \midrule
    1  & $0.000 \pm 0.008$ mm  & 1, 2    & 1  \\ 
    2  & $196.479 \pm 0.082$ mm & 3, 5, $4^{*}$ & 2, $4^{*}$  \\ 
    3  & $196.479 \pm 0.082$ mm & 6, 7, $8^{*}$ & 3, $4^{*}$  \\ 
    \bottomrule
\end{tabular}
\caption{Specifications of the three interferometers \textemdash the differential optical pathlength between the interfering beams $\Delta L$, the 
\gls{qpr} responsible for interferometric measurement and phasemeter reading the \glspl{qpr} for further processing.}
\label{tab:ifo_specs}
\end{table}

The \gls{zifo} optical bench and surrounding test setup is shown in Figure \ref{fig:instrument design}. The optical bench is placed within a metallic thermal shield to increase the thermal stability of the demonstrator. The combined setup is placed inside the ERIOS vacuum tank located at the Laboratoire d'Astrophysique de Marseille, providing a high-vacuum environment and stable thermal conditions for performance measurement of the \gls{zifo}. The dedicated ultra-stable laser system, phase measurement system, and data acquisition systems are located in the vicinity of the vacuum tank.

The \gls{zifo} houses three \gls{ifo} on a Zerodur\footnote{a ceramic glass with an ultra-low mean \gls{cte} = $0 \pm 0.007 \times10^{-6} \text{K}^{-1}$ for temperatures $\in [0,50] ^{\circ}\text{C}$\cite{Hartmann:2021}} baseplate. The optical scheme of the interferometers is shown in \fref{fig:instrument design}, displaying one equal-arm and two unequal-arm interferometers, representative of the split interferometry technique described in Section \ref{sec:introduction}. Two Nd:YAG lasers feed the bench with two infrared beams of wavelength \qty{1.064}{\micro\meter}.

The equal-arm interferometer IFO$_1$, measured via its two output ports \gls{qpr}$_1$ and \gls{qpr}$_2$, is used as a phase reference between the two laser beams injected onto the bench. The equal arm length ensures cancellation of \gls{lfn}, resulting in a measurement that is only affected by the interferometric noise of interest. This interferometer is also sensitive to thermoelastic noise originating from the optical fibers that guide the laser light from outside the vacuum tank towards the bench and are thus subjected to the thermal environment and gradients in air.

The other two interferometers, IFO$_2$ (\gls{qpr}$_3$, \gls{qpr}$_4$, \gls{qpr}$_5$) and IFO$_3$ (\gls{qpr}$_6$, \gls{qpr}$_7$, \gls{qpr}$_8$), of unequal arm length allow to estimate the thermoelastic gradients within the optical bench itself, which cause expansion and contraction of the bench and local variations in the refractive index of the optics, leading to pointing variations. These interferometers also allow for the quantification of the residual laser frequency noise due to the unequal arm lengths of the lasers in IFO$_2$ and IFO$_3$, as well as all other interferometric noise. Optical pathlength differences between the arms of the interferometers, with the corresponding \gls{qpr} and phasemeter unit numbers, can be found in Table \ref{tab:ifo_specs}.

The \glspl{qpr} measuring the beatnote were developed by the Artemis laboratory with support from APC and CNES, using commercial photodiodes made from InGaAs \cite{Hamamatsu:2021G6849Specs}\cite{bruhier}. Balanced detection is implemented in each of the three \glspl{ifo}; IFO$_1$ is equipped with a QPR at each of the two output ports, for redundancy, to reduce the \gls{rin} at the heterodyne frequency, 1f-RIN \cite{Stierlin1986ExcessnoiseSI},  and potential stray light, which are cancelled out due to the $\pi$ offset caused by an additional reflection between each port. IFO$_2$ and IFO$_3$ have three \glspl{qpr} each; one \gls{qpr} receives twice the optical power of the other two, in order to maintain the balanced power required for balanced detection. In addition to the eight \glspl{qpr} used for interferometric measurements, four \gls{sepr} monitor the laser power and allow for on-bench optical power stabilization.

The Nd:YAG lasers used in this setup are limited by their frequency noise with the corresponding noise \gls{asd} of  \begin{equation}
S_\nu^{1/2}(f) = 10^{4} \times \left( \frac{1~\text{Hz}}{f} \right)~\frac{\text{Hz}}{\sqrt{\text{Hz}}} .
\end{equation}
If not reduced, this frequency noise would translate into an optical pathlength noise of roughly \qty{7}{\pico\meter\per\sqrt{\hertz}} at 1 Hz (with a 1/f dependency of the laser noise spectrum, yielding much higher noise at lower frequencies), this value was derived using 
\begin{equation}
S^{1/2}_{\Delta \nu}= S^{1/2}_{\nu} \frac{\lambda}{c} \Delta \text{L},
\label{eq:asd}
\end{equation}
where $\lambda$ is the laser wavelength and $\Delta \text{L}$ is the difference in optical path of the two lasers (as defined in \tref{tab:ifo_specs}). Consequently, the lasers used on the \gls{zifo} are stabilized using a two-step servo-loop control system, developed at \gls{lte}, Paris. This system is designed to maintain frequency noise with an \gls{asd} below the LISA requirement of $S^{1/2}_{\nu} = \qty{30}{\hertz\per\sqrt{\hertz}}$ in the [\qty{1}{\milli\hertz}, \qty{10}{\hertz}] range. The frequency stabilization is performed in two steps. First, the two lasers are assigned as primary (L$_1$) and secondary (L$_2$), with the secondary maintaining a phase lock with a constant frequency offset from 5 to \qty{25}{\mega\hertz} relative to the primary. The secondary laser thus inherits the same frequency noise as the primary. In the second step, the primary laser is phase-locked to an ultra-stable \qty{1596}{\nano\meter} telecom laser after undergoing frequency tripling using nonlinear optics. This telecom laser is frequency-locked to a hyperfine absorption line of molecular iodine ${}^{127}\text{I}_2$ \cite{iodine_cell}. This two-step process results in a residual\footnote{\gls{lfn} post-beam recombination} \gls{lfn} noise \gls{asd}, $S^{1/2}_{\Delta \nu}$, of \qty{8.5e-7}{\pico\meter\per\sqrt{\hertz}} and \qty{2.1e-2}{\pico\meter\per\sqrt{\hertz}} in the equal and unequal-arm interferometers, respectively.
A detailed description of the laser systems used in the \gls{zifo} can be found in \cite{Mehlman:2023ics}.

 
The voltage signals of the \gls{qpr} quadrants are digitized by a phasemeter via an \gls{adc}. A pilot tone is added to the signal prior to digitization to correct for sampling jitter noise from the \gls{adc}.
Each phasemeter has eight channels, measuring signals from the quadrants of two \glspl{qpr}. The individual quadrant signals are sampled at \qty{47.5}{\hertz} and processed both as a sum and as a difference within each \gls{qpr}. The phasemeter provides the instantaneous frequency (averaged over the quadrants), which is later integrated, after subtraction of the mean, to obtain the phase evolution.
The summed signal gives the mean phase measured by the \gls{qpr}, corresponding to the longitudinal pathlength of the beams, while the differential signal between quadrants provides information on \gls{ttl} noise via a technique called \gls{dws}. This noise arises from cross-coupling of the phase measurement with lateral and angular jitters between the interferometric beams.
The phasemeters used within the \gls{zifo} experiment were designed at the \gls{aei} in Hannover \cite{Barke:2014} and further upgraded at the APC and CEA\footnote{Commissariat à l'énergie atomique et aux énergies alternatives}, Paris, to meet \gls{zifo}'s requirements for \gls{adc} jitter correction. Details on the development of the phase measurement system can be found in \cite{LISA:2024hlh}. The data from the phasemeters is processed and analyzed via dedicated scripts that also compute \gls{asd} values for the monitored parameters. This data acquisition and analysis is explored in the next section.
\section{Data acquisition and processing}
\label{sec:acquisition}

The \gls{zifo} optical bench test campaign was conducted in the ERIOS vacuum chamber at the \gls{lam}. This vacuum chamber maintains a pressure of \qty{1e-5}{\milli\bar} and an ambient temperature variation of $\pm$\qty{2}{\kelvin}. Operating in vacuum allows for the isolation of the optical path from residual particles in the surrounding medium and enables higher thermal stability. Prior to conducting phase-stability tests, several verification tests were performed to validate the nominal functioning point of the system and to demonstrate the pointing stability of the beam on the center of the \glspl{qpr} in air and in vacuum. The optical pathlength stability was evaluated over long-duration runs, where the data acquisition lasted for approximately 15 hours. During these runs, the two injected laser beams operate in a balanced power configuration of \qty{2}{\milli\watt\per beam}. Each run is performed at a constant heterodyne and pilot tone frequency in the $f \in [5,25]$ \unit{\mega\hertz} range, and the data is stored in a measurement file. We discuss three stability runs, given in \tref{tab:runs}, that are representative of the behavior explored within the scope of this paper.
\begin{table}[h]
\centering
\renewcommand{\arraystretch}{1.3} 
\setlength{\tabcolsep}{10pt} 
\begin{tabular}{@{}c c c@{}} 
    \toprule
    \textbf{Run} & \textbf{Beatnote $f_h$ [MHz]} & \textbf{Pilot tone $f_p$ [MHz]} \\ 
    \midrule
    A  & $16.001$  & $22.001$ \\ 
    B  & $16.001$ & $18.001$ \\
    C  & $5.001$  & $7.001$  \\  
    \bottomrule
\end{tabular}
\caption{Frequency specifications for three long-duration acquisition runs.}
\label{tab:runs}
\end{table}
The phasemeter data from each measurement file is processed in several stages via a python data analysis pipeline. An initial pass through the data computes the average pilot tone frequency and phase differences between channels and identifies any data packet losses. This preprocessing stage also applies an internal jitter correction to remove timing noise via the addition of a fixed frequency pilot tone. The results are then used to estimate timing offsets within each phasemeter and between different phasemeters using a noise cross-correlation and interpolation procedure to achieve sub-sample accuracy.

A delay filter is further applied to correct for timing offsets between channels.
After that, the time series is low-pass filtered to prevent aliasing and to smooth high-frequency noise. The corrected data is then calibrated: frequency, amplitude, and differential quadrant phase measurements from different channels are rescaled to physical units and combined to form averaged and differential quantities. The total beatnote frequency of one \gls{qpr} is defined as the mean of the frequencies of each quadrant as follows:
\begin{equation}
\nu_{i} = \frac{1}{4}\left(\nu_i^a + \nu_i^b + \nu_i^c + \nu_i^d\right),
\end{equation}
where $i \in [1,8]$ is the index of the \gls{qpr}. 
Additional derived parameters, such as differential wavefront sensing (DWS) signals, are reconstructed using the quadrant phase measurements\footnote{The phasemeter measures amplitude and frequency of the beatnote along with the quadrant phase difference $\phi_{i}^{a-b}$, $\phi_{i}^{b-c}$ $\phi_{i}^{c-d}$ and  $\phi_{i}^{d-a}$  within a photoreceiver $i$ allowing for retrieval of DWS} $\phi_{i}^{j}$ of a photoreceiver $i$ with quadrant $ j \in [\text{a,b,c,d}]$ yielding the two following DWS channels,
\begin{align}
\text{DWS}_{i}^{\text{ud}} &= \frac{\phi_i^a + \phi_i^b}{2} - \frac{\phi_i^c + \phi_i^d}{2}, \label{eq:dws_ud} \\
\text{DWS}_{i}^{\text{lr}} &= \frac{\phi_i^a + \phi_i^c}{2} - \frac{\phi_i^b + \phi_i^d}{2}, \label{eq:dws_lr} 
\end{align}
where \eref{eq:dws_ud} is the differential phase along the lateral axis (up-down) and \eref{eq:dws_lr} along the vertical axis (left-right).
The calibrated phasemeter data is further downsampled to \qty{4.7}{\hertz} in order to preserve amplitude and phase information while substantially reducing data volume. Prior to downsampling, a Kaiser-windowed low-pass filter with an attenuation of \qty{240}{\decibel}, designed to suppress spectral components above the \qty{2.3}{\hertz} Nyquist frequency is applied to the data, ensuring that decimation does not introduce aliasing. The downsampling has no consequence in terms of data loss because the \gls{lisa} bandwidth is bounded at \qty{1}{\hertz}.

The time-synchronised and downsampled frequency data from the four phasemeters is combined to create a single data channel for each interferometer. The signal for each interferometer is defined as the mean value of each port of the interferometer, as follows:
\begin{equation}
\text{IFO}_k = \frac{\sum_{i \in I_k} \nu_{i}}{|I_k|}, 
\end{equation}
where $I_k$ denotes the set of \glspl{qpr} corresponding to IFO$_k$ given in \tref{tab:ifo_specs}. 

Finally, the differential frequency time series is computed between pairs of interferometers, IFO$_{i/j}$ = IFO$_i$ - IFO$_j$, to isolate the beatnote frequency from common-mode noises, yielding the following combinations: IFO$_{2/1}$, IFO$_{3/2}$, and IFO$_{3/1}$. Subtracting the frequency measured by one interferometer relative to another allows the suppression of common and correlated noise, resulting in a noise reduction of five orders of magnitude between the mean and differential values. 
The next section deals with the analysis of the phasemeter data—to verify the pathlength stability and identify noise sources on the bench. Plots are presented in terms of the \gls{asd} of the optical phase in units of \unit{cycles\per\sqrt{\hertz}}, where one cycle in phase is equivalent to one wavelength in displacement. 
\section{Measurement performance and noise assessment}
\label{sec:performanceandnoisereduction}
\begin{figure}
    \centering
    \includegraphics[width=\linewidth]{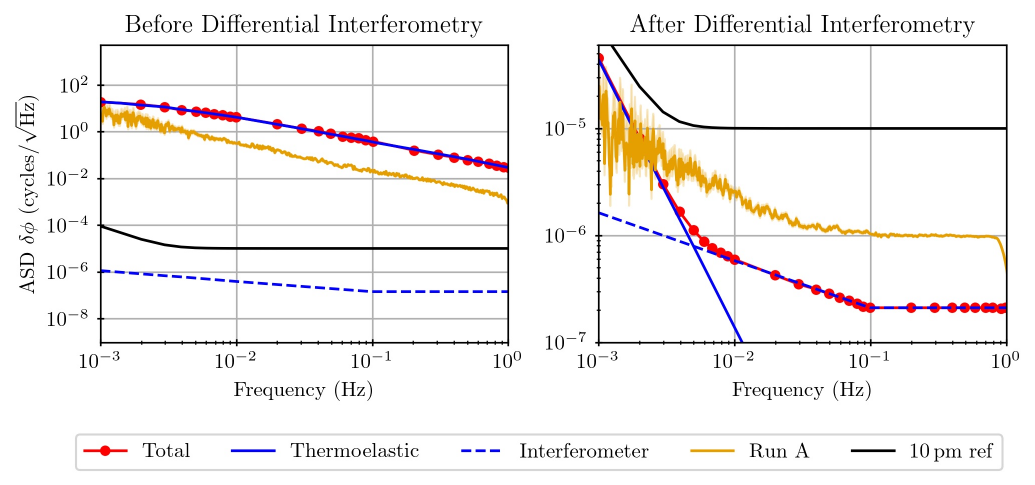}
    \caption{Thermoelastic and interferometer noise models for IFO$_1$ (left) and IFO$_{2/1}$  (right) differential interferometer measurement. Phase noise in interferometers during Run A is overlaid to enable comparison between model and measurement. The \qty{10}{\pico\meter} reference accounts for a relaxation factor of ${1}/{f^2}$ below \qty{3}{\milli\hertz}, as is LISA, the noise floor below this frequency is determined by test-mass acceleration noise \cite{LISA:2024hlh} and not optical metrology noise.}
    \label{fig:beforeandafterdifferential}
\end{figure}
The optical pathlength stability on the test-bench is driven by three main parameters: interferometer noise, tilt-to-length coupling, and thermoelastic noise. Secondary noise sources include those from vibrations and sample timing errors. The current performance model of the \gls{zifo} only includes contributions from the instrumental noise and thermoelastic noise, excluding those from \gls{ttl}, vibrations, clock noise and straylight\footnote{Note that straylight on the \gls{zifo} was later investigated in a separate study~\cite{Nardello:2020fio}.} for simplicity sake. Details on the individual sources and the performance model are explored in \ref{app:PerformanceModel}.


The instrumental noise model, shown in dashed blue in \fref{fig:beforeandafterdifferential} includes contributions from chain noise, shot noise, relative intensity noise and residual \gls{lfn}, the leading contribution attributed to phase measurement noise. The influence of thermoelastic noise (solid blue) depends on multiple factors like the thermal expansion of the bench and the optical components along the path of the laser beam, with the largest contributions coming from the optical fibre thermoelastic due to length and thermal gradient mismatch between the optic fibres of the two lasers. The noise models are overlaid with total phase noise in IFO$_1$ and differential combination IFO$_{2/1}$ from the validation Run A(yellow). Figure \ref{fig:beforeandafterdifferential} (left) shows how the dominant thermoelastic noise (solid blue) contribution to the total noise (dotted red) violates the \qty{10}{\pico\meter\per\sqrt{\hertz}} specification (black) by over seven orders of magnitude. Comparing the total noise (red) with the data from stability Run A (yellow), we see that phase before differential interferometry depicts noise lower by approximately an order of magnitude than that predicted by the noise model. This suggests that the noise on the optical bench, in particular thermoelastic noise, was lower on the bench than as upper limit estimated in the performance model. This idea is revisited below.

Optical fibre and some optical component thermoelastic noise are common-mode noise \textit{i.e.} correlated between \gls{qpr}s, and can be suppressed by conducting further analysis with the differential interferometer channels 
IFO$_{i/j}$ constructed in \sref{sec:acquisition}. However, because each beam in the unequal-arm interferometers follows a partially independent optical path, some thermoelastic noise remains unsuppressed as a residual. The resulting total phase noise (red) is illustrated in \fref{fig:beforeandafterdifferential} (right), dominated by the residual thermoelastic contribution at frequencies lower than \qty{5}{\milli\hertz} and by phasemeter-dominant instrumental noise above \qty{10}{\milli\hertz}. The residual thermoelastic noise contains contributions from inhomogeneous expansion of the Zerodur bench which does not get removed during differential calculation because the phase noise contribution has opposite signs in the \gls{qpr}s of IFO$_2$ and IFO$_3$. The slope of residual thermoelastic noise model after differential calculation, shown in \fref{fig:beforeandafterdifferential}, forms an upper limit of the thermal noise and is not representative to that on the bench, because thermoelastic noise contributions in the model are estimated using temperature data obtained from the thermal sensors on the metal thermal shield and not on the Zerodur bench itself. This claim is verified by the observed differential phase in Run A, which has lower noise at frequencies below \qty{3}{\milli\hertz} than that predicted by the noise model. At frequencies higher than \qty{3}{\milli\hertz}, the differential phase in Run A is swamped by noise contributions not accounted for in the performance model. 

The stability measurements presented in this section are influenced exclusively by non-common mode contributions of instrumental noise, \gls{ttl}, straylight, vibration noise and residual bench thermoelastic noise. It thus serves as an upper limit characterization of noise contributions to the optical bench that are not included in the performance model. We discuss the best pathlength measurements in \sref{sec:stability}. Noise anomalies identified during the campaign are discussed in \sref{sec:noisyqprs}. In \sref{sec:ttl}, we isolate and suggest mitigation techniques for contributions from correlations in \gls{dws} channels like \gls{ttl} coupling.

\subsection{Pathlength Stability}
\label{sec:stability}


\begin{figure}
    \centering
     \caption*{Run A: $f_h=16.001\,\mathrm{MHz}$, $f_p = 22.001\,\mathrm{MHz}$}
    \includegraphics{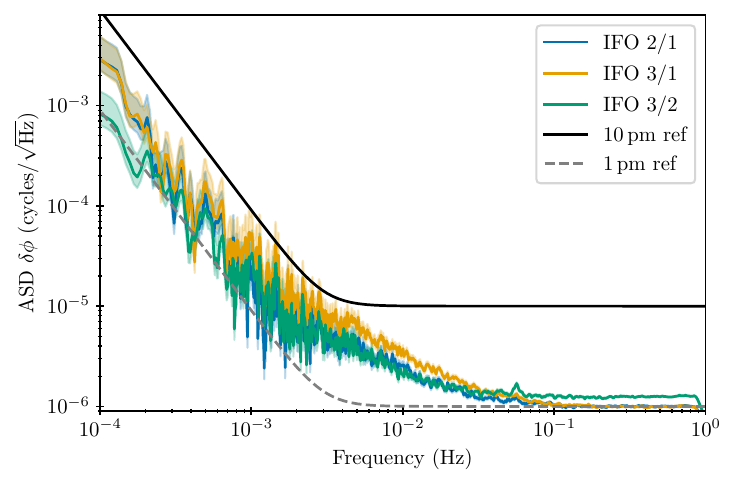}
    \caption{The amplitude spectral density of differential phase \textit{i.e.} the optical pathlength variation, for cyclic combinations of the three interferometers IFO$_{2/1}$ (blue), IFO$_{3/1}$ (orange), IFO$_{3/2}$ (green).}
    \label{fig:stability}
\end{figure}


The primary objective of the \gls{zifo} campaign was the demonstration of picometer-level noise conformity on the optical bench above the \qty{10}{\milli\hertz} band. We present the results in \fref{fig:stability}, where we plot differential phase noise between interferometers in terms of their amplitude spectral density for data acquisition Run A, before further noise analysis. Run A exhibits near perfect conformity of phase noise, all three differential combinations within the \qty{10}{\pico\meter\per\sqrt{\Hz}} reference curve (black). At higher frequencies, the noise in the optical system goes down to the \qty{1}{\pico\meter\per\sqrt{\Hz}} (dashed grey) reference curve. 
Figure \ref{fig:stability} depicts error bars of large peaks of IFO$_{3/1}$ that touch the requirement curve. Note that this issue is minor, but nevertheless resolved post-removal of correlated \gls{dws} noise using a technique explored in \ref{sec:ttl}, plotted in Appendix \ref{fig:Run537DWScomparison}.

Run B and C exhibit a residual noise that exceeds the \qty{10}{\pico\meter\per\sqrt{\hertz}} reference curve. The origin and removal of this noise is explored in the next section.

\subsection{Noisy \glspl{qpr}}
\label{sec:noisyqprs}

\begin{figure}[h]
\centering
\caption*{Run B: $f_h=16.001\,\mathrm{MHz}$, $f_p = 18.001\,\mathrm{MHz}$}
\includegraphics{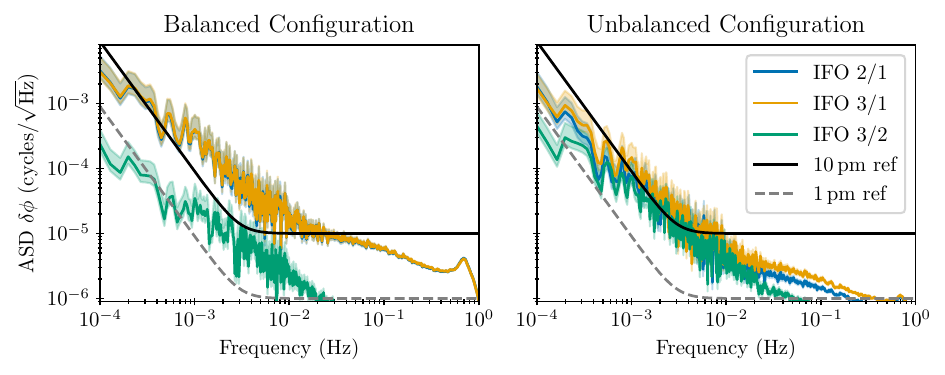} 
\caption{The amplitude spectral density of differential phase in balanced configuration and unbalanced configuration for Run B. 
    }
\label{fig:comparison547}
\end{figure}

\begin{figure}[h]
\centering
\caption*{Run C: $f_h=5.001\,\mathrm{MHz}$, $f_p = 7.001\,\mathrm{MHz}$}
\includegraphics{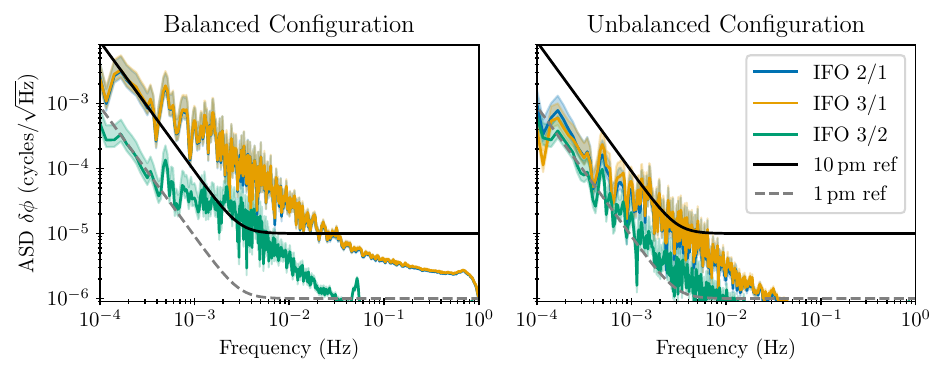} 
\caption{The amplitude spectral density of differential phase in balanced configuration and unbalanced configuration for Run C. 
    }
\label{fig:comparison557}
\end{figure}
Unlike Run A, the long-duration stability Runs B and C exhibit a differential phase \gls{asd} raised by one order of magnitude in the differential configurations of IFO$_{2/1}$ and IFO$_{3/1}$, violating the \qty{10}{\pico\meter\per\sqrt{\hertz}} reference curve. This phase noise is presented in \fref{fig:comparison547} and \ref{fig:comparison557}, where we compare the differential phase, for two configurations of the \gls{zifo}. On the left, the \gls{zifo} is in the conventional, balanced configuration; where all readout of all eight \gls{qpr} are used in the final phase measurement. This is compared with that of the unbalanced configuration (right); where we omit the readout of two \gls{qpr}s; QPR$_4$ and QPR$_8$. 

The source of this inconsistency was narrowed down to a common mode noise that uniquely affects QPR$_4$ and QPR$_8$ \textemdash associated with IFO$_2$ and IFO$_3$ respectively \textemdash due to the larger phase noise associated with the phasemeter measuring these \gls{qpr}; PM 4. This anomalous phase noise in PM 4 can be attributed to the non-uniform distribution of the pilot tone signals used for\gls{adc} jitter correction.
This pilot tone is generated then split between the four phasemeters \textit{i.e.} the channels of phasemeters 1,2 and 3 are on individual frequency distribution boxes but the ones of PM 4 are split between two boxes.  Furthermore, PM 4 is based on \gls{fmc} 107 instead of \gls{fmc}-108 like the other three phasemeters (with one using a 50 MHz clock and the others using a 100 MHz clock). This common mode phase noise from PM $4$ would get suppressed in IFO$_{3/2}$ (that employs both \glspl{qpr} 4 and 8) but not on IFO$_{2/1}$ and IFO$_{3/1}$ (that only use one per combination). In \fref{fig:comparison547} we observe that when the \gls{zifo} is in unbalanced configuration, the noise floor of IFO$_{2/1}$ and IFO$_{3/1}$ reduces by an order of magnitude, the one of IFO$_{3/2}$ gets slightly noisier. It is suspected that the cause of the marginal increase in IFO$_{3/2}$ is owed to the retro-reflected straylight on the optical bench \footnote{Identified in~\cite{Nardello2024StrayLight}.} or to 1f-\gls{rin} at the beatnote frequency of the laser. 
In balanced detection when all the \gls{qpr} are running and optical powers are balanced at the output ports of the \gls{ifo}, we observe that part of the noise due to stray light instability in the heterodyne beatnote cancels out, as expected, in balanced detection interferometry, with stray light entering by the port opposite to the nominal port.

The variation in differential phase noise \gls{asd} between IFO$_{2/1,3/1}$ and IFO$_{3/2}$ fluctuates by $2.8 - 5 \times 10^{-3}$ over the long-duration runs (also see \fref{fig:553comparison}). This suggests the presence of other noises and environmental dependencies in addition to that of the noisy \gls{qpr} in the runs. Therefore, it follows that while the removal of QPR$_4$ and QPR$_8$ from the dataset reduces the phase noise added by \gls{pm} 4, the overall phase noise in IFO$_{2/1}$ and IFO$_{3/1}$ post \gls{qpr} removal varies between different runs. Presently, no correlation has been deduced between the run parameters and the magnitude of phase noise discrepancy between the differential interferometric combinations. However, we anticipate that phase noise arising from \gls{pm} 4 can be reduced or even completely omitted by splitting all the channels of all four phasemeters between the pilot tone frequency distribution boxes. This would in turn result in the same noise affecting all phasemeters, making it common mode noise, which is canceled in the differential measurement values. This should be verified in further work.

We observe that noise disparity still remains despite removing contributions from \gls{qpr}$_4$ and \gls{qpr}$_8$. This effect is more pronounced in Run C as illustrated in \fref{fig:comparison557} under unbalanced detection. We discuss a technique to reduce this phase noise disparity in the next section. 
\subsection{Removal of Correlated Noise}
\label{sec:ttl}

Data acquisition runs for long-term phase stability test exhibited an increased phase noise in combinations with a difference in pathlength (IFO$_{2/1,3/1}$) even after subtraction of noisy \gls{qpr}s. Section \ref{sec:architecture} introduced how the technique of \gls{dws} is used to quantify relative angular mismatch between the interfering beams on the \gls{qpr}. Any correlated noise in the \gls{dws} channels would get subtracted in the equal arms of IFO$_{1}$, but not in the unequal arm interferometers IFO$_{2}$ and IFO$_{3}$. Therefore, any \gls{ttl}-like correlated noise present in the unequal interferometer would remain after the equal-unequal differential calculation and increase the phase noise. One possible source of noise in the \gls{dws} channels could be \gls{ttl} coupling, originating from the jitter of the laser beams.

Focused work on suppression of \gls{ttl} in LISA by linear modelling of coefficients for total jitter per degree of freedom in single-links and \gls{tdi} variables can be found in \cite{Paczkowski:2022nrt,Houba:2022wni,Hartig:2024ojo}.
This single-link model was extended to include \gls{ttl} in individual interferometers in \cite{Wanner:2024eoa}, which will be particularly relevant if \gls{ttl} calibration manoeuvres are performed on LISA \cite{Houba:2022vwq}.   
In this section, we propose an ad-hoc method to detect and subtract correlations between \gls{dws} channels and differential pathlength readout with possible contributions from \gls{ttl}. Note that this method would result in the removal of any correlated noise like correlations of thermal noise between beam pointing and bench expansion, and not just \gls{ttl}. 

Each differential interferometric combination post removal of noisy \gls{qpr}s given in \sref{sec:noisyqprs} involves four \gls{qpr}s, resulting in eight \gls{dws} channels. High correlation between these \gls{dws} channels indicates common-mode coupling in the participating interferometers. We propose a two-step noise reduction strategy. First, we perform dimensionality reduction on the correlated \gls{dws} channels. Second, we employ linearity assumptions to determine the coupling constants that scale the correlated jitter into the phase measurements. The assumption of linearity has been deemed sufficient based on the fit and subtraction of \gls{ttl} in the \gls{lpf} mission~\cite{Armano:2023fmc}.

The problem of dimensionality reduction is addressed using \gls{pca} \cite{pearson1901pca,Hotelling:1933PCA}. This technique creates a new basis of \textit{principal components} consisting of orthonormal vectors chosen along axes that maximize data variance. This basis is created by using singular value decomposition of the cross-correlation matrix of the \gls{dws} channels. The resulting eigenvalues are rearranged in descending order, and summed to create principal components in decreasing order of variance. Using \gls{pca}, we transform the \textit{basis} of ten interdependent \gls{dws} channels to  $m$ linearly-independent components that explain $99.5\%$ of the original variance. 

Assuming a linear relation between our reduced basis and the differential phase $\delta \phi$, we express the measured phase difference as:
\begin{equation}
    \delta \phi = \delta\phi_0 + \sum_i^m\epsilon_i \gamma_i + n_{ij},
    \label{eqn:totalphasenoise}
\end{equation}
where $\delta\phi_0$ denotes the noise-free differential phase, $\gamma_i,\forall i \in m$ represents $m$ basis components of \gls{dws} measurements and $n_{ij}$ encapsulates other instrument noise sources.
\begin{figure}
    \centering
\includegraphics{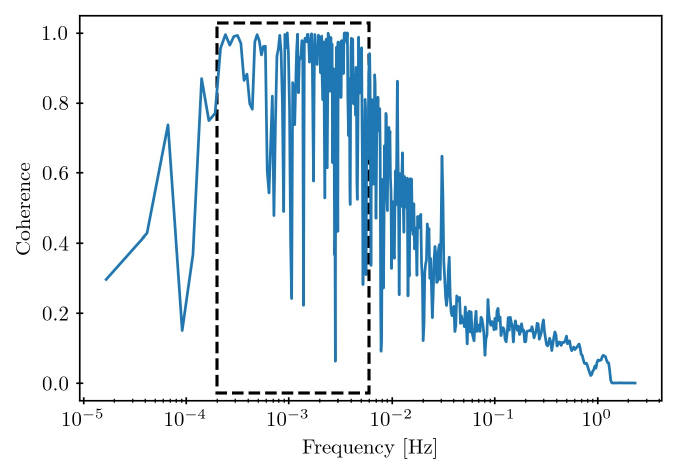}
    \caption{Coherence between the phase in IFO$_{2/1}$ and the first principal component for Run C, the high coherence region marked by the dashed rectangle.}
    \label{fig:coherence}
\end{figure}
In our analysis, we find a maximum of $m=2$ components are highly correlated with the differential phase measurement. Physically, this means that maximally two independent noise sources are common to the \gls{dws} readout of different photoreceivers, including contributions from beam jitter.
The proportionality constant for each basis component $\epsilon_i$ is estimated using:
\begin{equation}
\epsilon_i
= \int_{f_c} \frac{\left\langle \phi(f) \gamma_i^{*}(f) \right\rangle}{\left\langle \left| \gamma_i(f) \right|^{2} \right\rangle}
\label{eqn:propconstant}
\end{equation}
where $^*$ denotes complex conjugation and integration is performed over frequencies with high coherence $f_c$. A typical coherence plot between phase and basis components for $\epsilon$ estimation is depicted in \fref{fig:coherence}. Estimation of $\epsilon$ allows the subtraction of the maximum \gls{dws} variance from the differential phase in the frequency range of \qty{0.2}{\milli\hertz} to \qty{5}{\milli\hertz} where the coherence is significant.
\begin{figure}[h]
\centering
\caption*{Run C: $f_h=5.001\,\mathrm{MHz}$, $f_p = 7.001\,\mathrm{MHz}$}
\includegraphics{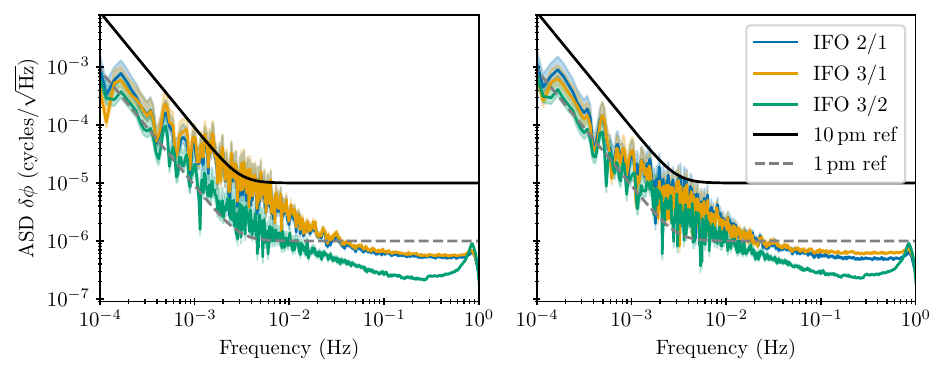} 
\caption{The amplitude spectral density of differential phase before(left) and after(right) DWS correlation reduction.}
\label{fig:comparison557pca}
\end{figure}

The differential phase for a long duration run before(left) and after(right) removal of correlated noise are illustrated in \fref{fig:comparison557pca}. Here, we only subtract the first principal component $i=1$ (57\% of \gls{dws} variance) 
\begin{equation}
   \delta \hat{\phi} = \delta \phi \, - \, \epsilon_1\gamma_1 =  \delta\phi_0 + \sum_{i=2}^m\epsilon_i \gamma_i + n_{ij}
\end{equation}
to prevent inadvertent noise injection through overcorrection\footnote{A similar phase noise addition from over-subtraction was seen in the \gls{lpf} mission \cite{Armano:2023fmc}, corrected by test-mass realignment post March 2016.}. Post subtraction, the phase noise of IFO$_{3/1}$ and IFO$_{2/1}$ shows a significant reduction between \qty{0.5}{\milli\hertz} to \qty{10}{\milli\hertz}. The effect on the noise floor above \qty{10}{\milli\hertz} is negligible.  

\gls{pca} analysis was conducted on acquisition Run A and Run B with difference initial noise levels presented in \fref{fig:Run537DWScomparison} and \fref{fig:Run547DWScomparison} show smaller improvements in comparison to the one presented above for Run C. This technique thus has reproducibility limitations due to its dependence on initial noise level, which are strongly influenced by environmental testing conditions. Some of the acquisitions performed during the test campaign experienced noisier environment condition (acoustic, thermal..) which could explain why this technique improved noisier runs but could not be standardised to achieve improvement in all runs. Details on the environment anomalies are discussed in \sref{sec:environment}. It would be interesting to compare correlation coefficients found using the \gls{dws} correlation method discussed here with the ones predicted by optical simulations and dedicated \gls{ttl} experiments in follow up studies.



\subsection{Test set up temperature stability}
\begin{figure}[h]
\centering
\includegraphics[width=0.9\linewidth]{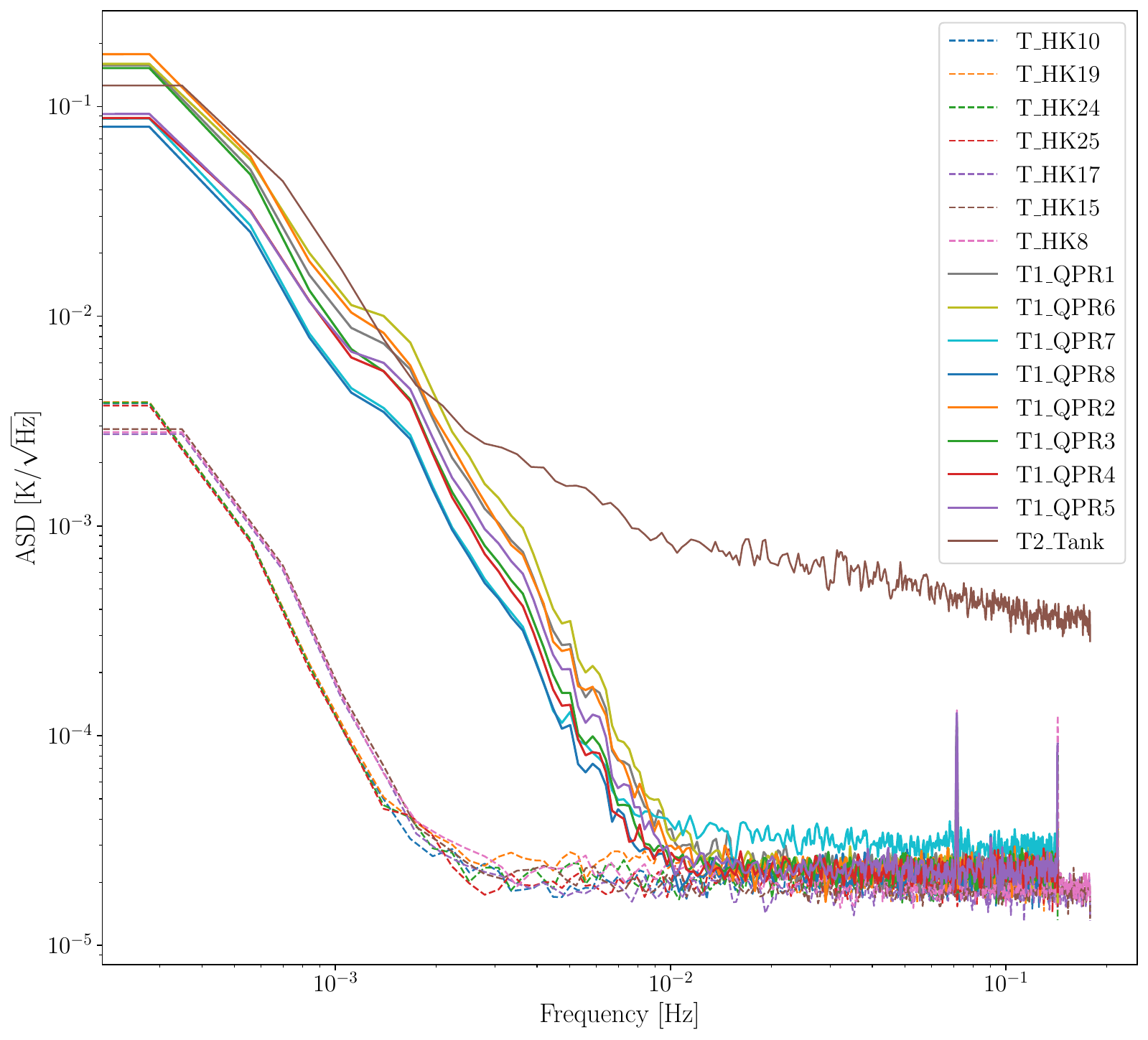} 
\caption{The amplitude spectral density of the temperature measurement for Run C for different temperature sensors inside the photo-receivers, on the thermal shield and on the vacuum tank.}
\label{fig:temp_stab}
\end{figure}

The test setup is equipped with multiple thermal sensors to verify nominal operation of the equipment and to characterize the thermal environment and achievable stability. Two types of sensors were employed: NTC 50 kOhm thermistors (negative temperature coefficient, with a nominal resistance of 50 kOhm) and PT 10 kOhm platinum temperature sensors. 
Temperature sensors were installed inside each photo-receiver (T\_QPR), at various locations on the thermal shield (T\_HK), and on the outer wall of the vacuum chamber (T\_Tank). The amplitude spectral density of the temperature measured by the NTC sensors is shown in Figure~\ref{fig:temp_stab}. Above 1~mHz, the stability is limited by the acquisition system, which constrains the performance to approximately \(\sim 20~\mu\mathrm{K}/\sqrt{\mathrm{Hz}}\).
For both the thermal shield sensors and the QPR sensors, the same behavior is observed: the noise floor is reached in the high-frequency part of the spectrum, and below a cut-off frequency the noise level increases. The cut-off frequency for the thermal shield sensors is around 2~mHz and appears to be dominated by spectral leakage from the DC component and the finite acquisition time. For the QPR sensors, the cut-off frequency is higher, at approximately 10~mHz, reflecting the fact that these sensors are located within active dissipative components. 
No thermal sensors were placed on the Zerodur baseplate; the stability measured on the thermal shield is the closest proxy for the one on the baseplate, and it is likely not as good. This implies that the thermal stability on the Zerodur baseplate is at worst \(\sim 20~\mu\mathrm{K}/\sqrt{\mathrm{Hz}}\).

\section{Test environment}
\label{sec:environment}
As briefly mentioned in \sref{sec:ttl}, the environment at \gls{lam} had some uncharacterised features that disturbed the test campaign.$\,$These included \gls{emc} disturbances, primarily originating by the start-up phase of the primary and secondary pumping systems required to maintain tank vacuum. Measurement systems like dataloggers can also cause conductive \gls{emc} disturbances that can propagate to the various data acquisition systems and contaminate the data. 

The acoustic environment was also degraded by the cryogenic pump that enabled to reach a high vacuum, those disturbances were causing vibrations on the optical fiber connecting the laser benches to the vacuum tank, leading to large glitches in the beatnote frequency measured by the phasemeter. For this reason, a weaker vacuum was implemented only using the primary pump and turbo molecular pump that allowed to reach a \qty{1e-5}{\milli\bar} vacuum.

The seismic noise and vibration environment at \gls{lam} was of outstanding quality, leading to the selection of this facility for the test campaign. The optical table inside the vacuum chamber is mechanically isolated from both the chamber and the building, being supported by a \qty{100}{T} seismic block.

Additionally, the thermal environment at LAM exhibited temperature fluctuations of $\pm$ 1,5$^{\circ}$C due to the very large clean room volume to regulate. Exposure of the the optical fibres and their connections to these temperature cycles cause significant polarisation rotations, leading to optical power variations that were as large as double the standard value. This lead to loss of laser locking between the two laser benches. Eventually the problem was mitigated by thermally insulating the fibres for the long-duration runs that needed to maintain laser phase locking for over 24 hours. The testing campaign thus stresses on a full characterization of environmental parameters like electromagnetic disturbances, thermal gradients and stability or acoustic environment.

\section{Conclusion}
\label{sec:conclusion}

The phase noise measurement illustrated in \fref{fig:stability} demonstrates the success of \gls{zifo}'s primary objective; reduction of noise on the optical bench to reach and maintain the \gls{lisa} requirement of picometer-level optical pathlength stability at frequencies ranging from \qty{1}{\milli\hertz} to \qty{1}{\hertz}. Furthermore, the development of the \gls{zifo} enabled the testing of key technologies required in the future test campaigns of the interferometric core of LISA; updating the phase measurement system, validating \qty{1}{\milli\meter} diameter laser injection and integration of optical contacting on Zerodur. 

We discuss sources and post-processing noise reduction techniques in data acquisitions that fail to meet the picometer requirement. Dominant noise in the equal-unequal combinations of IFO$_{2/1}$ and IFO$_{3/1}$ was identified as a common-mode noise originating from the non-uniform pilot tone frequency generation in the four phasemeters. This was locally isolated from affected runs by removing \gls{qpr}$_4$ and \gls{qpr}$_8$ from interferometric averaging. Future validation experiments should verify if the noise is eliminated by either splitting quadrant channels of all \gls{qpr}s on different frequency distribution boxes or ensuring all phasemeters exhibit the same synchronisation performance. 

We identified the presence of correlations between lateral or angular tilt of the beam and the pathlength, visible as correlations between \gls{dws} channels. The source of these correlations could be from \gls{ttl} or from the fluctuating thermal profile of the testing environment. This noise was removed using \gls{pca}-driven dimensionality reduction and subsequent subtraction using a linearised model. As predicted, the noise removal technique only reduced the phase noise in equal-unequal combinations and either added noise or didn't affect the unequal-unequal one.

Despite all the operational challenges faced during the \gls{zifo} campaign, all objectives were achieved and some even surpassed in the case of maintaining pathlength stability. Furthermore, the development and verification of the demonstrator highlights the cohesion of the LISA-France community and its collaboration with industry partners to realize a common deliverable. The successful \gls{zifo} campaign paves the way for the \gls{bsim} being developed in the next phase of the technology development for \gls{lisa}.

\section*{Acknowledgements}
The authors gratefully acknowledge the support of \gls{cnes} for both financial and technical assistance throughout the project duration, including the Ph.D. funding of MV. We also thank the Albert Einstein Institute for their contribution to the design of the phase measurement system. 
\clearpage

\appendix

\section{Performance Model}
\label{app:PerformanceModel}

As introduced in \Sref{sec:performanceandnoisereduction}, the \gls{zifo} performance budget considers contributions from instrumental and thermal noise. It excludes contributions from \gls{ttl}, vibrations and clock noise. In this section, we examine the origin and contribution of each noise source in brief. Values of referenced constants can be found in \tref{tab:noise_contributions}.

\paragraph{Instrumental Noise:}
\label{app:introduction}

Instrumental noise in each interferometer I, \gls{eq} arm (IFO$_1$) and \gls{uneq} arm (IFO$_{2,3}$), and their responsible \gls{qpr}$_a,\forall a\in[1,8]$, is a quadratic sum of the \gls{asd}s of chain noise, phase measurement noise, shot noise, \gls{rin}${_1f}$, \gls{rin}$_{2f}$ and residual frequency noise. Note that noise from laser light scattering was excluded at time of performance model generation due to lack of noise estimate. Subsequent studies on straylight contributions to the \gls{zifo} can be found in \cite{Nardello:2020fio}. 

\begin{equation}
\tilde{s}_{\mathrm{I}, \mathrm{QPR}_a}=\sqrt{\tilde{s}_{\mathrm{QPR}_a, \text {Chain}}^2+\tilde{s}_{\mathrm{PMS}, \mathrm{pm}}^2+\tilde{s}_{\mathrm{QPR}_a, \text {shot}}^2+\tilde{s}_{\mathrm{RIN}_{1\mathrm{f}}, \mathrm{I} }^2+\tilde{s}_{\mathrm{RIN}_ {2\mathrm{f}}}^2+\tilde{s}_{\text {Res } f, \mathrm{I}}^2}\,. 
\end{equation}

Chain noise derives from the coupling of the noise in the front-end electronics of the photoreceivers into the phase noise and thus inversely proportional to laser power, is given by~\cite{Barke:2014lsa}
\begin{equation}
\tilde{s}_{\mathrm{QPR}_a, \text { Chain }}=\frac{\lambda}{2 \pi} \frac{\tilde{s}_{\text {C}} \sqrt{N_{\text {Seg}}}}{R_{P D} \sqrt{2 \eta P_{1\mathrm{QPR}_a} P_{2\mathrm{QPR}_a}}}
\end{equation}
where

\begin{tabular}{ll}
$\lambda$          & wavelength of laser [\si{\nano\meter}] \\
$\tilde{s}_{\text {C}}$     & \gls{qpr} input current noise density for \qty{20}{\mega\hertz} beatnote [\si{\pico\ampere\per\sqrt{\hertz}}] \\
$N_{\text {Seg}}$          & number of \gls{qpr} segments \\
$R_{P D}$          & responsivity of InGaAs [\si{\ampere\per\watt}]\\
$P_{1(2)\mathrm{QPR}_a}$ & power of laser 1 (laser 2) on \gls{qpr}$_a$ [\si{\milli\watt}]\\
$\eta$          & heterodyne efficiency. \\ \\ 
\end{tabular}

Chain noise is higher in the \gls{qpr}s $4,5,7\,\text{and}\,8$ as they receive a lower laser power $P_{1\mathrm{QPR}}$ of \qty{0.41}{\milli\watt} compared to the \num{0.83}-\qty{0.92}{\milli\watt} received by \gls{qpr} $1,2,3\,\text{and}\,6$. The phase measurement noise can be attributed to the output noise from the front-end electronics of the photoreceivers. Leading contribution to interferometer noise comes from the phasemeter, with a maximum value of \qty{1.16}{\pico\meter\per\sqrt{\hertz}} at \qty{1}{\milli\hertz} going down to \qty{0.15}{\pico\meter\per\sqrt{\hertz}} above \qty{100}{\milli\hertz}. Akin to chain noise, shot noise is also inversely proportional to laser power. It is a quantum noise effect related to the discreteness of photons and electrons and sets
a fundamental limit to the optical intensity noise in measurement set-ups using photoreceivers such that
\begin{equation}
\tilde{s}_{\mathrm{QPR}_a, \text { shot }}=\frac{\lambda}{2 \pi} \sqrt{\frac{q_e\left(P_{1\mathrm{QPR}_a}+P_{2\mathrm{QPR}_a}\right)}{R_{\mathrm{PD}} \,\eta \,P_{1\mathrm{QPR}_a} P_{2\mathrm{QPR}_a}}}
\end{equation}
where $q_e$ is the charge of an electron. A larger contribution of shot noise is expected onboard LISA than the one measured on the \gls{zifo} as the power of the two interfering beams on the \gls{zifo} is of the same order of magnitude. \gls{rin} refers to the fluctuations in optical power of the laser at the heterodyne beatnote frequency \gls{rin}${_{1f}}$ or at twice the beatnote \gls{rin}${_{2f}}$, and appears as an amplitude modulation on the laser beam. \gls{rin}${_{1f}}$ depends on the power of the laser impinging on the relevant beam splitter for each interferometer $I$
\begin{equation}
\tilde{s}_{\mathrm{RIN} 1 \mathrm{f}, \mathrm{I}}=\frac{\lambda}{2 \pi} \sqrt{\frac{P_{1\mathrm{BS}_I}^2+P_{2\mathrm{BS}_I}^2}{2\, \eta\, P_{1\mathrm{BS}_I} \tilde{P}_{2\mathrm{BS}_I}}} \tilde{s}_{\mathrm{RIN}}\left(1-\epsilon_{\text {balance }}\right)
\end{equation}
where

\begin{tabular}{ll}
$\tilde{P}_{1(2)\mathrm{BS}_I}$          & laser 1 (2) power on beam splitter $I$ [\si{\milli\watt}], \\
$\tilde{s}_{\mathrm{RIN}}$          & laser \gls{rin} [\si{unit\per\sqrt{\hertz}}], \\
$\epsilon_{\text {balance}}$     & beam amplitude ratio on two sides of beam splitter. \\ \\
\end{tabular}

The \gls{lpf} mission showed that \gls{rin} can be removed by conducting measurements in balanced detection mode, because the noise at either port of the beam splitter is equal. When balanced detection is disturbed e.g. due to noisy \gls{qpr}s as explored in \ref{sec:noisyqprs}, \gls{rin} contributes to phase noise. Furthermore, residual \gls{rin} will remain if the coating of the beam splitters is imperfect, or due to mismatch of the end-to-end electronics of the \gls{qpr}s. Contrarily, \gls{rin}${_{2f}}$ cannot be completely removed via balanced detection. Its contribution to phase noise in the \gls{zifo} is independent of laser power and is given by the simple relation 
\begin{equation}
    \tilde{s}_{\mathrm{RIN} 2 \mathrm{f}} = \frac{\lambda}{2\pi} \frac{\tilde{s}_{\mathrm{RIN}}}{2}.
\end{equation}
The residual frequency noise originates of the imperfect cancellation of laser frequency noise. It depends on the mean difference in arm-length of the interferometer 
\begin{equation}
\tilde{s}_{\text {Res }f, \mathrm{ I}}=\frac{\lambda \Delta L_{\mathrm{I}}}{c} \tilde{s}_{\nu}
\end{equation}
where 

\begin{tabular}{ll}
$\Delta L_{\mathrm{I}}$          & armlength difference in interferometer I [\si{\milli\meter}], \\
$\tilde{s}_{\nu}$          & laser frequency noise [\si{\hertz\per\sqrt{\hertz}}], \\
$c$     & speed of light [\si{\meter\per\second}]. \\ \\
\end{tabular}
\paragraph{Thermoelastic Noise:} Sources of thermal fluctuations that couple to the pathlength on the \gls{zifo} include the thermal expansion of the optical bench Zerodur base plate and the thermo-elastic effect in the optical fibres that supply the laser beam to the injectors. Thermo-elastic noise of individual optical components beyond the beam-splitter of each interferometer is a common-mode noise and thus removed during differential interferometer calculation. Expansion of the optical bench itself is a also common mode noise that gets cancelled. However, thermal noise due to bench expansion in the differential modes of each interferometer depends on the armlength difference in the combining beams $\Delta L_{\mathrm{EQ}}$ or $\Delta L_{\mathrm{UNEQ}}$. It is given by 
\begin{equation}
\tilde{s}_{\text {Therm }, \mathrm{OB}, \text { Zer }}=\alpha_{\mathrm{CTE}, \text { Zer }} \tilde{s}_{\mathrm{T_{OB}}} \Delta L_{\mathrm{I}}
\end{equation}
where 

\begin{tabular}{ll}
$\alpha_{\mathrm{CTE}, \text { Zer }}$          & thermal expansion coefficient Zerodur [\si{\per\kelvin}], \\
$\tilde{s}_{\mathrm{T_{OB}}}$          & allocated local temperature stability [\si{\micro\kelvin\per\sqrt{\hertz}}]. \\ \\
\end{tabular}

Optical fibres made of fused silica provide the laser from the generation bench in air to the injectors of the \gls{zifo} kept in vacuum. Fibre thermoelastic noise originates from length mismatch between the fibres of the two laser and any thermal gradients induced between the two fibres. These contributions are correlated, and thus the total fibre thermoelastic noise can be expressed as 
\begin{equation}
\begin{array}{c}
\tilde{s}_{\text {Therm,Fiber }}=\left(n_{\mathrm{Si}} \alpha_{\mathrm{CTE}, \mathrm{Si}}+\frac{\mathrm{d} n_{\mathrm{Si}}}{\mathrm{~d} T}\right) \\
\times\left[\tilde{s}_{\mathrm{T}_{\text {air }}}\left(L_{\text {Fiber,air }}+\Delta L_{\text {Fiber,air }}\right)+\tilde{s}_{\mathrm{T}_{\text {vac }}}\left(L_{\text {Fiber,vac }}+\Delta L_{\text {Fiber,vac }}\right)\right]
\end{array}
\end{equation}
where 

\begin{tabular}{ll}
$n_{\mathrm{Si}}$          & fused silica refractive index, \\
$\alpha_{\mathrm{CTE}, \mathrm{Si}}$          & fused silica thermal expansion coefficient [\si{\per\kelvin}], \\ 
$\frac{\mathrm{d} n_{\mathrm{Si}}}{\mathrm{~d} T}$          & fused silica refractive index temperature coefficient [\si{\per\kelvin}], \\
$\tilde{s}_{\mathrm{T}_{\text {air(vac)}}}$          & environment stability in air (vacuum) at \gls{lam} [\si{\micro\kelvin\per\sqrt{\hertz}}], \\
$L_{\text {Fiber,air(vac) }}$          & nominal fibre length in air ( vacuum) [\si{\milli\meter}], \\
$\Delta L_{\text {Fiber,air(vac) }}$ & allocated difference between two fibres in air (vacuum) [\si{\milli\meter}].  \\ \\
\end{tabular}

Noise from the fibre remains the dominant source of noise resulting in a maximum total thermoelastic noise of \qty{1.85e7}{\pico\meter\per\sqrt{\hertz}} prior to differential interferometer calculation.

\paragraph{Vibration Noise:} Seismological measurements due to acceleration \gls{asd} due to vertical vibrations, conducted at the \gls{apc} laboratory on an measurement table mounted on shock absorbers. The vibration noise \gls{asd}, combined with flexion simulation on a \qty{5}{\centi\meter} thick Zerodur plate is calculated as \qty{1.58e-3}{\pico\meter\per\sqrt{\hertz}} using the formula
\begin{equation}
\tilde{s}_{\mathrm{vib}, \mathrm{APC}}=N_{\mathrm{oc}} \,D_{\mathrm{Zer}} \,\xi_{\mathrm{Zer}}\, \times 10^{\frac{\tilde{S}_{\mathrm{vib}, \mathrm{APC}}}{20}}
\end{equation}
where 

\begin{tabular}{ll}
$N_{\mathrm{oc}}$          & allocated maximal number of optical components in beam path \\
$D_{\mathrm{Zer}}$          & Zerodur optical bench allocated thickness [\si{\milli\meter}], \\ 
$\xi_{\mathrm{Zer}}$          & Zerodur optical bench simulated flexion [\si{\micro\radian\per\meter\per\second\squared}], \\
$\tilde{S}_{\mathrm{vib}, \mathrm{APC}}$          & vibration acceleration \gls{psd}  [\si{\decibel\meter\per\second\squared\per\hertz}]. \\ \\
\end{tabular}

This value is a minor contribution compared to other noise sources. Although equivalent measurements were not performed for the ERIOS chamber at \gls{lam}, its specified optical bench stability of $\leq 10^{-7} g$ indicates that vibration noise was a negligible contributor during \gls{zifo} data acquisition.

\paragraph{Tilt-to-length Noise:} Angular misalignments between interferometric beams couple to the interferometric readout, contributing to the phase noise in the form of tilt-to-length coupling. Here we discuss contributing mechanisms detailed in~\cite{Schuster:2017rzo} that could result in \gls{ttl} noise on the \gls{zifo}. Beams during recombination on photodiode have could be misaligned, resulting in a spurious measurement of phase differences. This misalignment could be constant or dynamic. Constant offset depends strongly on beam and detector geometry and is given by an odd polynomial of the tilt angle $\alpha$
\begin{equation}
    \Delta s_{\text{const. offset}} = \sum_n a_n \alpha^{2n+1}
\end{equation}
A dynamic offset induced between the rotation point and the photodiode leads to a dependant beam walk, given by
\begin{equation}
    \Delta s_{\text{dyn. offset}} = \sum_n a_n \alpha^{2n}
\end{equation}
In an experiment, constant and dynamic offset both contribute, and thus the resulting \gls{ttl} coupling can take any shape. Discrepancies in beam parameters like the wavefront curvatures of the interfering lasers in the photodiode plane result in a loss of contrast and change in overall phase. This coupling can be expressed as,
\begin{equation}
    \Delta s_{\text{beam params.}} = \sum_n a_n \alpha^{2n}
\end{equation}
and is usually dominated by the second order term.
If the two interfering beams are not fundamental Gaussian modes, the amplitude symmetry around the beam axis is broken, resulting in wavefront errors. This noise, while significant for the \gls{lisa} science interferometer, is a minor contributor for the \gls{zifo} as both beams are expected to be fundamental Gaussian modes. It can be expressed as
\begin{equation}
    \Delta s_{\text{wavefront}} = \sum_n a_n \alpha^{n}.
\end{equation}
If the segments of a QPD have different efficiencies or different shapes and defects resulting
in cross-talks, the balance of signals between segments is consequently disturbed and a
tilting wavefront induces a detector geometry TTL coupling~\cite{FernandezBarranco:2016}. This coupling can take any form,
\begin{equation}
    \Delta s_{\text{detector}} = \sum_n a_n \alpha^{n}.
\end{equation}
The contribution of each source can be assessed using optical simulations. Since each contribution is independent, a quadratic sum of each contribution would give an estimate of the non-geometrical \gls{ttl} on the \gls{zifo}.
\paragraph{Sample Timing Noise:} The timing noise associated with the mismatch of pilot tones between different phasemeters can contribute to phase noise. 
\begin{table}[h]
    \centering
    \renewcommand{\arraystretch}{1.3} 
    \setlength{\tabcolsep}{6pt} 
    \small 
    \begin{tabular}{l p{7.5cm} c}
        \toprule
        \textbf{Variable} & \textbf{Description} & \textbf{Value} \\
        \midrule
        $\lambda$          & Wavelength of the laser & 1064 \si{\nano\meter} \\
        $\tilde{s}_{\text {C}}$ & Photodiode current noise & 3.40 \si{\pico\ampere\per\sqrt{\hertz}} \\
        $N_{\text {Seg}}$ & QPR number of segments & 4 \\
        $R_{PD}$          & Photodiode responsivity & 0.69 \si{\ampere\per\watt} \\
        $L_{\text {ref}}$ & Reference coaxial cable length & 25 \si{\meter} \\
        $L_{\text {cab}}$ & Coaxial cable length on set-up & 10 \si{\meter} \\
        $A$               & Zerodur OB baseplate thickness & $-15.14$ \si{\decibel} \\
        $\eta$            & Heterodyne efficiency & 0.8 \\
        $\tilde{s}_{\mathrm{RIN}}$ & Relative intensity noise & $10^{-7}$\si{\/\per\sqrt{\hertz}} \\
        $\epsilon_{\text {balance}}$ & Relative matching of the amplitudes on either side of beam combiner & 0.9 \\
        $\Delta L_{\mathrm{EQ}}$ & Equal arm IFO pathlength difference & 1 \si{\milli\meter} \\
        $\Delta L_{\mathrm{UNEQ}}$ & Unequal arm IFO pathlength difference & 200 \si{\milli\meter} \\
        $\tilde{s}_{\nu}$ & Laser frequency noise & 30 \si{\hertz\per\sqrt{\hertz}} \\
        $\alpha_{\mathrm{CTE}, \text { Zer }}$ & Zerodur coefficient of thermal expansion & $2\times10^{-8}$ \si{\per\kelvin} \\
        $\tilde{s}_{\mathrm{T_{OB}}}$ & Local temperature stability & [\si{\micro\kelvin\per\sqrt{\hertz}}] \\
        $n_{\mathrm{Si}}$ & Refractive index of fused silica & 1.450 \\
        $\alpha_{\mathrm{CTE}, \mathrm{Si}}$ & Fused silica coefficient of thermal expansion & $0.5\times10^{-6}$ \si{\per\kelvin} \\ 
        $\frac{\mathrm{d} n_{\mathrm{Si}}}{\mathrm{~d} T}$ & Temperature dependence of fused silica index & $8.31\times10^{-6}$ [\si{\per\kelvin}] \\
        $\tilde{s}_{\mathrm{T}_{\text {air(vac)}}}$ & Environmental temperature stability in air/vacuum & [\si{\micro\kelvin\per\sqrt{\hertz}}] \\
        $L_{\text {Fiber,air}}$ & Fiber length in air & 3000 \si{\milli\meter} \\
        $L_{\text {Fiber,vac }}$ & Fiber length in vacuum & 5000 \si{\milli\meter} \\
        $\Delta L_{\text {Fiber,air }}$ & Fiber path difference in air & 50 \si{\milli\meter} \\
        $\Delta L_{\text {Fiber,vac }}$ & Fiber path difference in vacuum & 50 \si{\milli\meter} \\
        $N_{\mathrm{oc}}$ & Maximal number of optical components on beam path & 10 \\
        $D_{\mathrm{Zer}}$ & Zerodur optical bench baseplate thickness (allocated) & 50 \si{\milli\meter} \\ 
        $\xi_{\mathrm{Zer}}$ & Zerodur vibration coefficient & $10^{-2}$ \si{\micro\radian\per\meter\per\second\squared} \\
        $\tilde{S}_{\mathrm{vib}, \mathrm{APC}}$ & APC vibration noise & -130 \si{\decibel\meter\per\second\squared\per\hertz} \\
        \bottomrule
    \end{tabular}
    \caption{Constants and parameters used in the formulas, with explicit descriptions.}
    \label{tab:noise_contributions}
\end{table}



\section{Plots}
\label{app:plots}

\begin{figure} [h]
\centering
\includegraphics{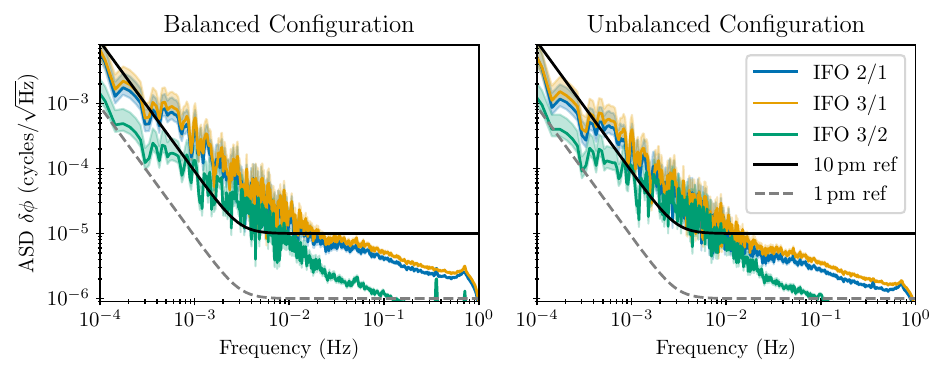} 
\caption{The amplitude spectral density of  differential phase in balanced configuration and unbalanced configuration for a long-duration stability test characterized by a heterodyne frequency of \qty{21.001}{\mega\Hz} and a pilot tone of \qty{23.001}{\mega\Hz}. 
    }
\label{fig:553comparison}
\end{figure}

\begin{figure}[h]
    \centering
    \caption*{Run A: $f_h=16.001\,\mathrm{MHz}$, $f_p = 22.001\,\mathrm{MHz}$}
    \includegraphics{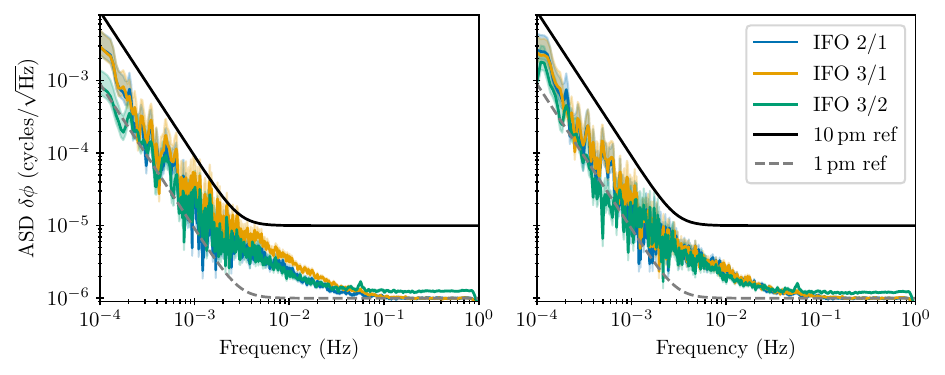}
    \caption{The amplitude spectral density of differential phase \textit{i.e.} the optical pathlength variation, for cyclic combinations of the three interferometers IFO$_{2/1}$ (blue), IFO$_{3/1}$ (orange), IFO$_{3/2}$ (green) prior(a) and post(b) removal of correlations in \gls{dws} channels. }
    \label{fig:Run537DWScomparison}
\end{figure}

\begin{figure}[h]
    \centering
    \caption*{Run B: $f_h=16.001\,\mathrm{MHz}$, $f_p = 18.001\,\mathrm{MHz}$}
    \includegraphics{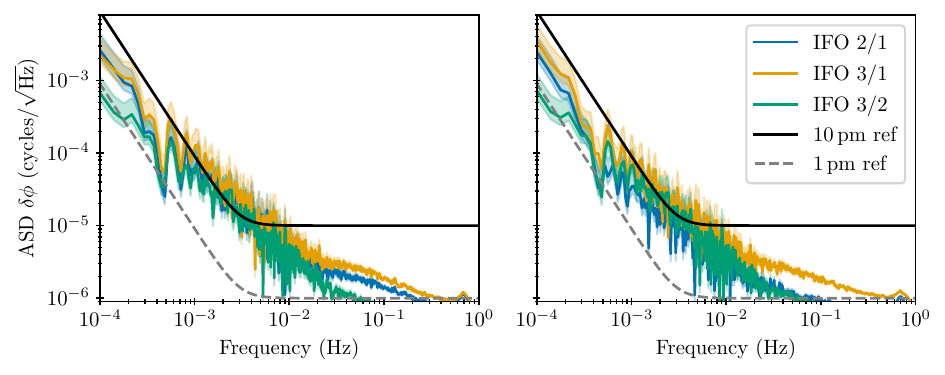}
    \caption{The amplitude spectral density of differential phase \textit{i.e.} the optical pathlength variation, for cyclic combinations of the three interferometers IFO$_{2/1}$ (blue), IFO$_{3/1}$ (orange), IFO$_{3/2}$ (green) prior(a) and post(b) removal of correlations in \gls{dws} channels. }
    \label{fig:Run547DWScomparison}
\end{figure}


\FloatBarrier
\section*{References}
\bibliographystyle{iopart-num}
\bibliography{refs}

\end{document}